\documentclass{webofc}
\usepackage[varg]{txfonts}   
\usepackage{hyperref}
\usepackage{url}
\usepackage{siunitx}
\usepackage{cleveref}
\usepackage{lineno}
\hypersetup{colorlinks=true,citecolor=blue,urlcolor=blue,linkcolor=blue}
\begin{document}
%
\title{LHCb Open Data Ntupling Service: On-demand production and publishing of custom LHCb Open Data}
%
%

\author{\firstname{Christine} \lastname{Aidala}\inst{1} \and
        \firstname{Dillon} \lastname{Fitzgerald}\inst{1} \and
        \firstname{Kai} \lastname{Habermann}\inst{2} \and
        \firstname{Ludwig} \lastname{Kramer}\inst{2} \and 
        \firstname{Adam} \lastname{Morris}\inst{3} \and
        \firstname{Sebastian} \lastname{Neubert}\inst{2} \and
        \firstname{Piet} \lastname{Nogga}\inst{2}\fnsep\thanks{\email{piet.nogga@cern.ch}} \and
        \firstname{Eduardo} \lastname{Rodrigues}\inst{4} \and
        \firstname{Marco} \lastname{Donadoni}\inst{3} \and
        \firstname{Daan} \lastname{Rosendal} \and
        \firstname{Tibor} \lastname{Šimko}\inst{3}
}

\institute{University of Michigan
\and
           University of Bonn
\and
           CERN
\and
            University of Liverpool
          }

\abstract{
The LHCb Ntupling Service enables on-demand production and publishing of LHCb Run 2 Open Data and aims at publishing them through the CERN Open Data Portal. It integrates the LHCb Ntuple Wizard to generate the configuration files that link custom user requests to the internal LHCb production system. User requests are managed through a streamlined workflow, from request creation to the final production of ROOT Ntuples, which can be downloaded directly from the Ntupling Service web interface, eliminating the need of running complex experiment-level software on the user side.
}
\maketitle
\section{Introduction}
\label{intro}
One of CERN's core values is its commitment to innovation, transparency, and integrity. This is especially reflected by its dedication to publish the data available across all LHC experiments. The CERN Open Data Policy~\cite{OpenDataPolicy} encourages the adoption of a consistent approach to publishing scientific data. In particular, this means that the data is made available at a stage which is accessible and useful to the global scientific community, and where latent, domain-specific knowledge is captured in relevant metadata, documentation and user interfaces. Here, CERN applies the FAIR standards~\cite{FAIR} that provide guidelines to improving \textbf{F}indability, \textbf{A}ccessibility, \textbf{I}nteroperability, and \textbf{R}euse of scientific data. This includes the release of reconstructed data~\cite{OpenDataPolicy} together with analysis-level experiment-specific software on the CERN Open Data Portal~\cite{OpenDataPortal}, which serves as a central access point for CERN Open Data across all LHC experiments. This follows the DPHEP definition for research-grade experimental data (Level 3 data)~\cite{DPHEP}, which is sufficient to perform complete analyses.

In accordance with the above commitments, the LHCb collaboration has agreed to publish the full data set of a data taking campaign after 10 years, and approximately 50\% of the data 5 years after the end of a running period leading to the publication of the full Run 1 data set in 2023~\cite{OpenDataPortal} amounting to
\SI{900}{\tera\byte} of reconstructed data. However, the structure of the released data requires users to operate the (open source) LHCb-specific software \texttt{DaVinci}~\cite{DaVinci,DaVinciGit} to obtain the typical columnar data output, the ROOT Ntuples~\cite{Ntuples}, used by the majority of analysts in High Energy Physics, which can be cumbersome. Furthermore, the significantly larger data sets recorded by LHCb during Run 2 and beyond means its open data release does not scale effectively due to data storage limitations. This motivates the development of a new release format for Run 2 Open Data and beyond.

Approaching 10 years after the completion of Run 2, a close collaboration between the CERN IT department and the LHCb collaboration has been formed resulting in a on-demand production and publishing system of LHCb Open Data utilizing the LHCb Ntuple Wizard~\cite{NtupleWizard} to generate configuration files for Analysis Productions, the internal LHCb production system~\cite{AnaProd}. The application presented in these proceedings, the LHCb Ntupling Service, introduces a novel approach to publishing open data, bridging the LHCb Ntuple Wizard with the CERN Open Data Portal. This allows users to create custom open data requests, making use of LHCb internal tools, and consequently download the Ntuples once the request has been completed. 

These proceedings are structured as follows: \Cref{Wizard} provides a brief overview of the LHCb Ntuple Wizard and how the application can create and customize open data requests. \Cref{NtuplingService} describes the LHCb Ntupling Service, highlighting in detail how it utilizes the LHCb Ntuple Wizard and facilitates the publication of open data through user submitted queries within the CERN Open Data Portal. \Cref{Conclusion} wraps up with a summary and considerations for future improvements.
\section{The LHCb Ntuple Wizard}
\label{Wizard}
\begin{figure}
    \centering
    \includegraphics[width=\linewidth]{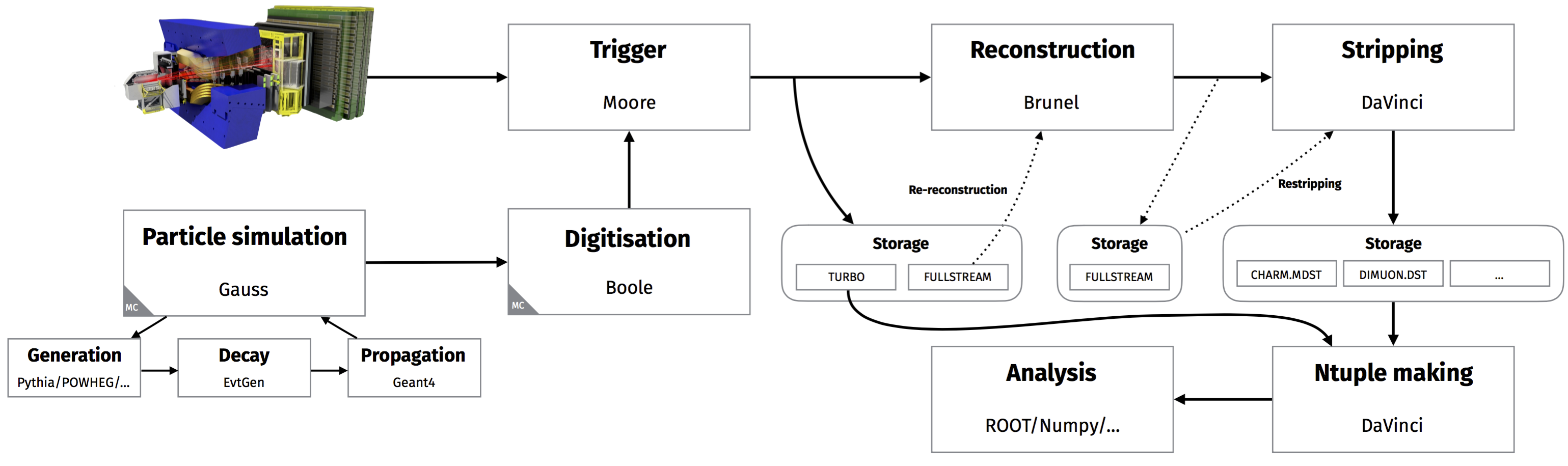}
    \caption{LHCb data flow in Run 2. The output of the Stripping step is made public.}
    \label{fig:data_flow}
\end{figure}

\begin{figure}
    \centering
    \includegraphics[width=0.8\linewidth]{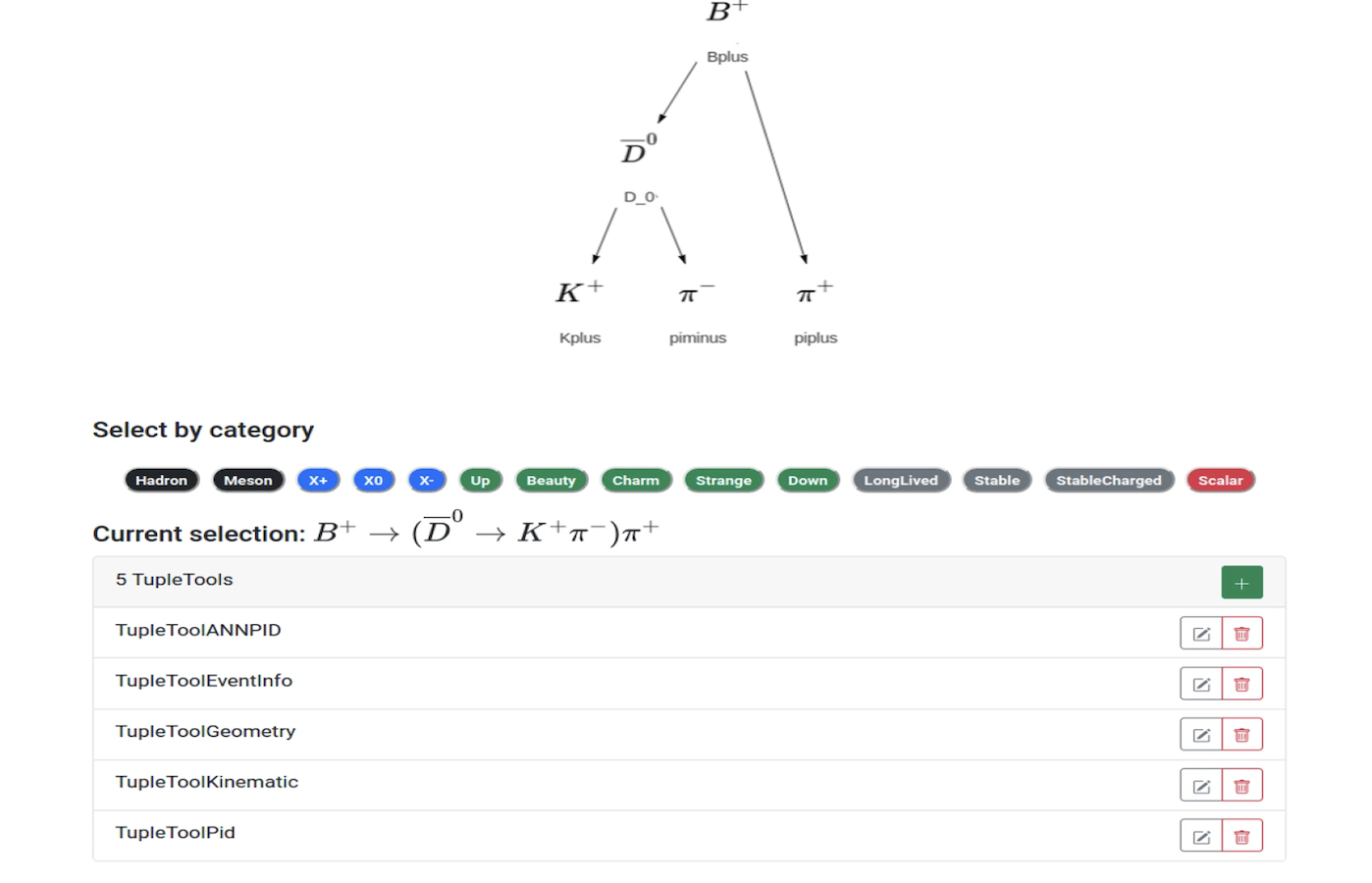}
    \caption{Snapshot of the LHCb Ntuple Wizard web interface, displaying the \textit{decay tree} and the customization of Ntuples via TupleTools. It is shown exemplary for the $B^+ \rightarrow \bar{D}^0 \pi^+$ decay~\cite{NtupleWizard}.}
    \label{fig:web_wizard}
\end{figure}

For the Run 2 open data release, the LHCb collaboration has committed to the reconstructed data described in~\Cref{intro} being the output of the preselection processing, called \textit{Stripping}~\cite{Stripping}. It is depicted in the LHCb Run 2 data flow in~\Cref{fig:data_flow}, and is a central component of the LHCb Data Processing \& Analysis (DPA) software project~\cite{DPA}.
In essence, the Stripping (processing stage) is designed to streamline the vast amounts of raw data collected from particle collisions into more physically meaningful and manageable subsets, called \textit{Streams}. They contain selections with similar physics signatures in order to be most useful for detailed scientific analyses. This is achieved by applying a series of algorithms using selection lines, which are essentially predefined sets of cuts or criteria, that further isolate the decay candidates of interest. Usually, they contain particles corresponding to a specific physics decay, for example a $B^+ \rightarrow \bar{D}^0\pi^+$ where the $\bar{D}^0$ decays into a $K^+\pi^-$ pair. The data structure of such physical processes is represented by a \textit{decay tree}, where all relevant variables to an analysis are stored in a ROOT Ntuple~\cite{Ntuples}. Both, the Stripping and the Ntupling are processed by the \texttt{DaVinci} application~\cite{DaVinci,DaVinciGit}. The application is configured using Python scripts, defining the sequence of algorithms used for the Ntupling. In particular, customizing the output Ntuples is possible by applying \textit{TupleTools} and \textit{LoKi functors}~\cite{Starterkit}, writing a well-defined set of variables to the Ntuple.
The visualization of a decay tree, as well as the customization of the Ntuple is integrated in the LHCb Ntuple Wizard, as depicted in~\Cref{fig:web_wizard}.

The LHCb Ntuple Wizard~\cite{NtupleWizard} is designed such that it extracts metadata and documentation about the available data and preselections of the Stripping selection lines and generates \texttt{DaVinci} configuration files based on user customizations on the desired decay chain and output variables. These configuration files are written such that it fits the architecture of the LHCb internal production system, Analysis Productions~\cite{AnaProd}, creating output Ntuples by running DIRAC~\cite{DIRAC1,DIRAC2} jobs on the CERN Worldwide LHC Computing Grid (WLCG). 

Several security considerations have been taken into account, since the generated code is eventually run on LHCb computing resources, with safeguards in place to prevent arbitrary code execution that could pose a risk to the infrastructure. Therefore, the configuration is written in YAML files as a pure data-structure format that is parsed internally and needs to be submitted to an LHCb Analysis Productions manager for processing, including a manual review of the request. More security considerations are discussed in~\Cref{sec:security}.

\begin{figure}
    \centering
    \includegraphics[width=\linewidth]{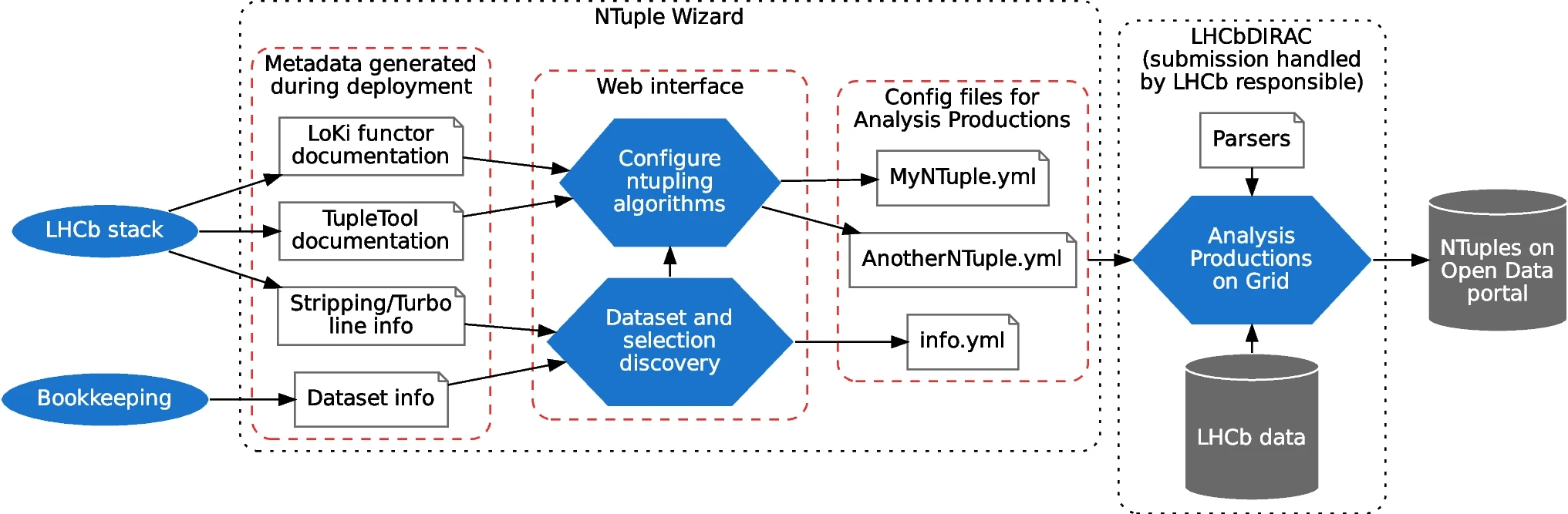}
    \caption{Architecture of the LHCb Ntuple Wizard~\cite{NtupleWizard}. The last step of publishing the output Ntuples to the CERN Open Data Portal is handled by the LHCb Ntupling Service, described in detail in~\Cref{NtuplingService}.}
    \label{fig:wizard}
\end{figure}

The workflow of the LHCb Ntuple Wizard is seen in~\Cref{fig:wizard}.
The Ntuple customization is handled through a user-friendly web interface which also serves as a guide through the configuration steps providing essential documentation for an open data user. This includes documentation for the Stripping and data set selection, and Ntuple configuration steps such as TupleTools and LoKi functors~\cite{Starterkit}.
\section{The LHCb Ntupling Service}
\label{NtuplingService}
\subsection{The LHCb Ntupling Service Architecture}
\begin{figure}
    \centering
    \includegraphics[width=\linewidth]{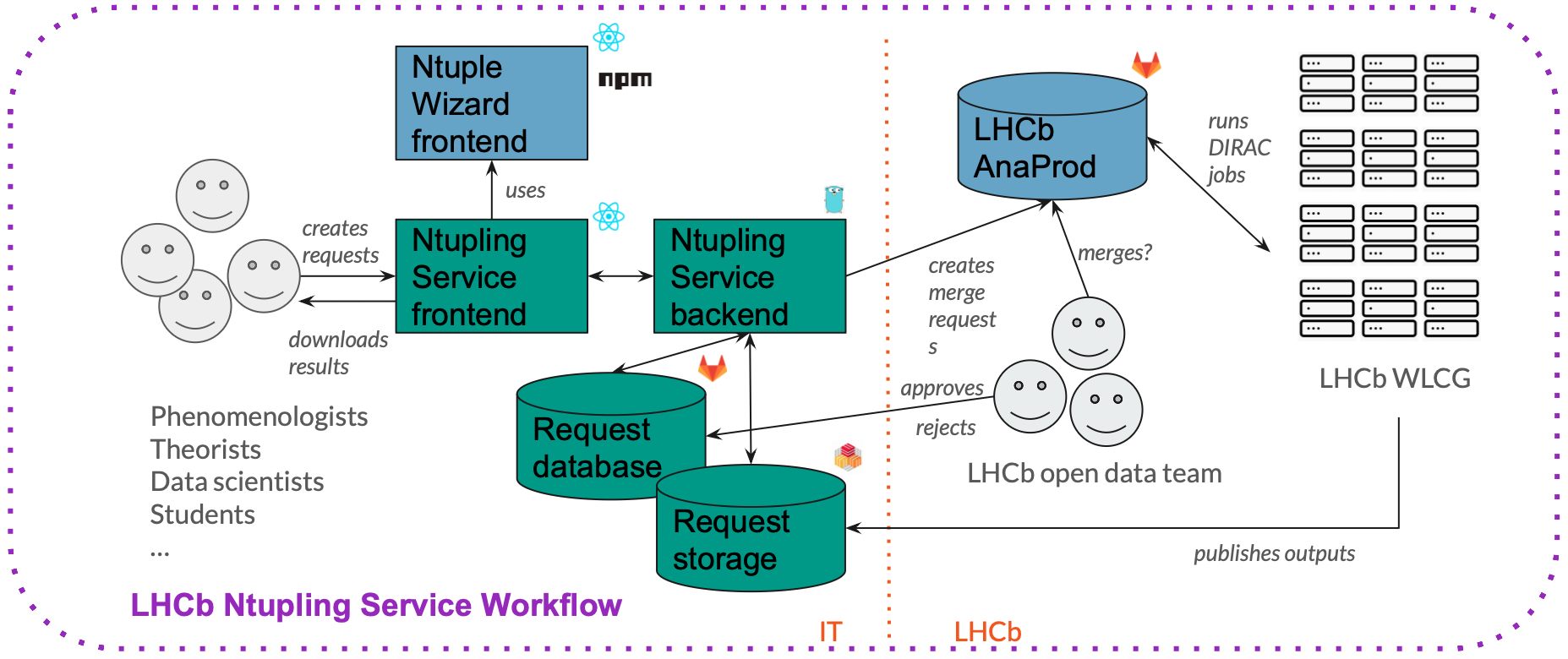}
    \caption{Architecture of the LHCb Ntupling Service. It consists of a frontend interface, embedding the LHCb Ntuple Wizard, a backend interface, which stores and propagates the generated configuration files by opening a production request to Analysis Productions, and a GitLab request repository, where the request is internally reviewed and processed. Furthermore, the user is able to download the output Ntuples directly from the web interface.}
    \label{fig:workflow_service}
\end{figure}

As mentioned in~\Cref{intro}, the larger data volumes recorded during Run 2 prompt a new release format for LHCb Open Data. The LHCb Ntuple Wizard offers a powerful and easy tool to request on-demand LHCb Ntuples. In order to efficiently streamline the process of creating an open data request to publishing the data on the CERN Open Data Portal, the LHCb Ntupling Service was developed. The architecture of the Ntupling Service is depicted in~\Cref{fig:workflow_service}. The following subsections will describe the components of the LHCb Ntupling Service and the workflow in detail.
\subsubsection{Web Interface}
\begin{figure}
    \centering
    \includegraphics[width=\linewidth]{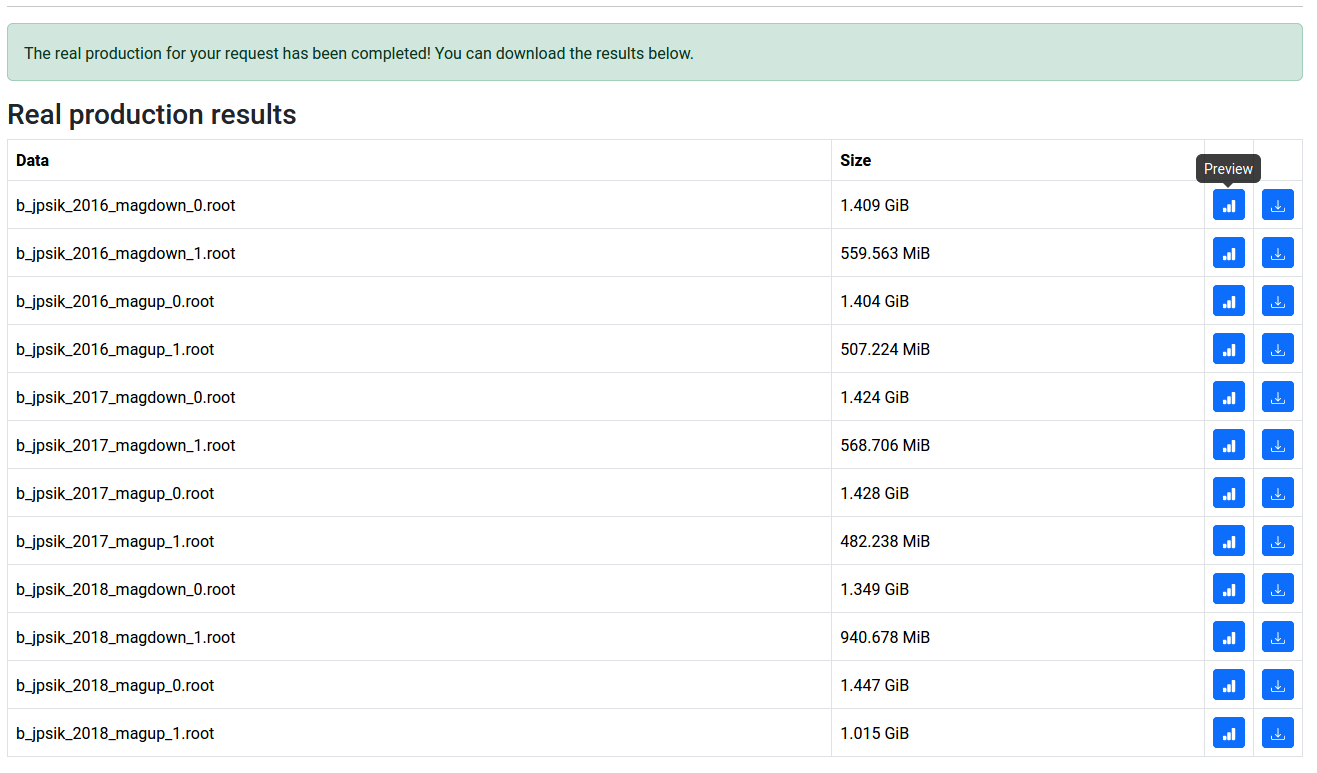}
    \caption{The CERN Open Data Portal hosts the output Ntuples, which users will be able to view and download on-demand.}
    \label{fig:download_data}
\end{figure}

The frontend component of the Ntupling Service is built on ReactJS~\cite{ReactJS}, embedding the LHCb Ntuple Wizard, where users can create custom Ntuple requests according to~\Cref{Wizard}. Here, it is also possible to view the status of the request or change it on-demand. Furthermore, there is a comment feature allowing the user to communicate with the LHCb Open Data team. Lastly, after the production has finished, the output Ntuples are propagated to the LHCb Ntupling Service web interface, where the user will be able to download the data as seen in~\Cref{fig:download_data}. The application requires user identification for safety purposes, which also includes guest access based on various identity providers such as Google, GitHub or an external email. Upon creating a request, the user is asked to create a profile, where, for example, the field of research must be inserted. This is interesting for the LHCb collaboration to further improve the application and keep track of common interests in the open data community.
\subsubsection{The LHCb Ntupling Service backend}
\begin{figure}
    \centering
    \includegraphics[width=0.9\linewidth]{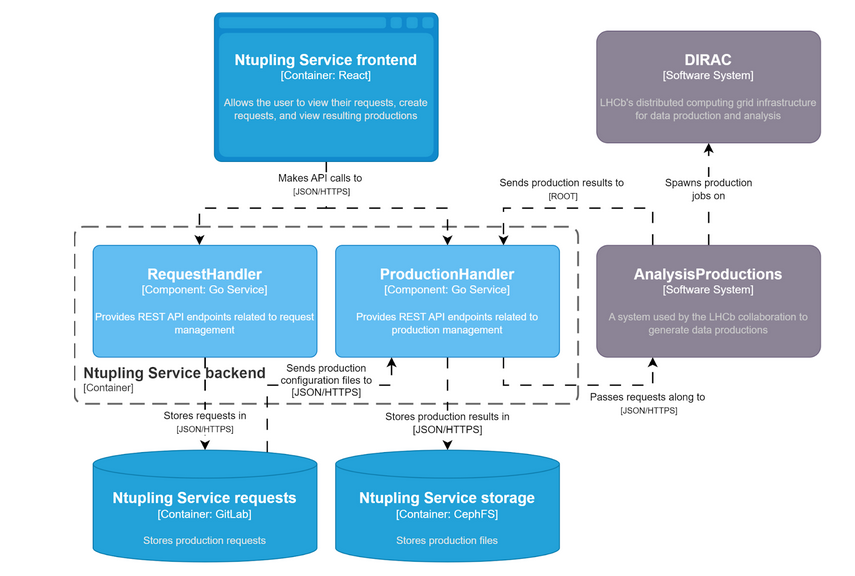}
    \caption{Component Diagram of the LHCb Ntupling Service backend. The application consists of two handlers, where the \textit{RequestHandler} and \textit{ProductionHandler} provide REST API endpoints related to the request management and production management, respectively.}
    \label{fig:service_backend}
\end{figure}

The backend of the Ntupling Service is built on Go~\cite{Go} and designed to manage both request and production processes efficiently. It includes a \textit{RequestHandler}, which is responsible for handling incoming user tasks such as creating new requests and managing them in an internal database. Additionally, the backend features a \textit{ProductionHandler}, which, after the approval of user requests, oversees production-related operations, including initiating test productions in Analysis Productions, storing results from both test and final productions, and transferring the resulting files back to the requester. A schematic view of the Ntupling Service backend implementation is illustrated in~\Cref{fig:service_backend}.
\subsubsection{The LHCb Ntupling Service Workflow}
Once a request is created on the user side, an API call is sent to the LHCb Ntupling Service backend, which will streamline the process of Ntuple production until the output Ntuples are produced and made available for user download.
First, the production configuration files are stored by the Ntupling Service and the associated user request is created as an issue in a GitLab repository. The system uses issues and corresponding merge requests to store user requests, where the source branch contains all relevant files. This enables the LHCb Open Data team to thoroughly review the production files and subsequently decide whether to accept or reject the request based on its compatibility with the LHCb computing resources.

\begin{figure}
    \centering
    \includegraphics[width=0.8\linewidth]{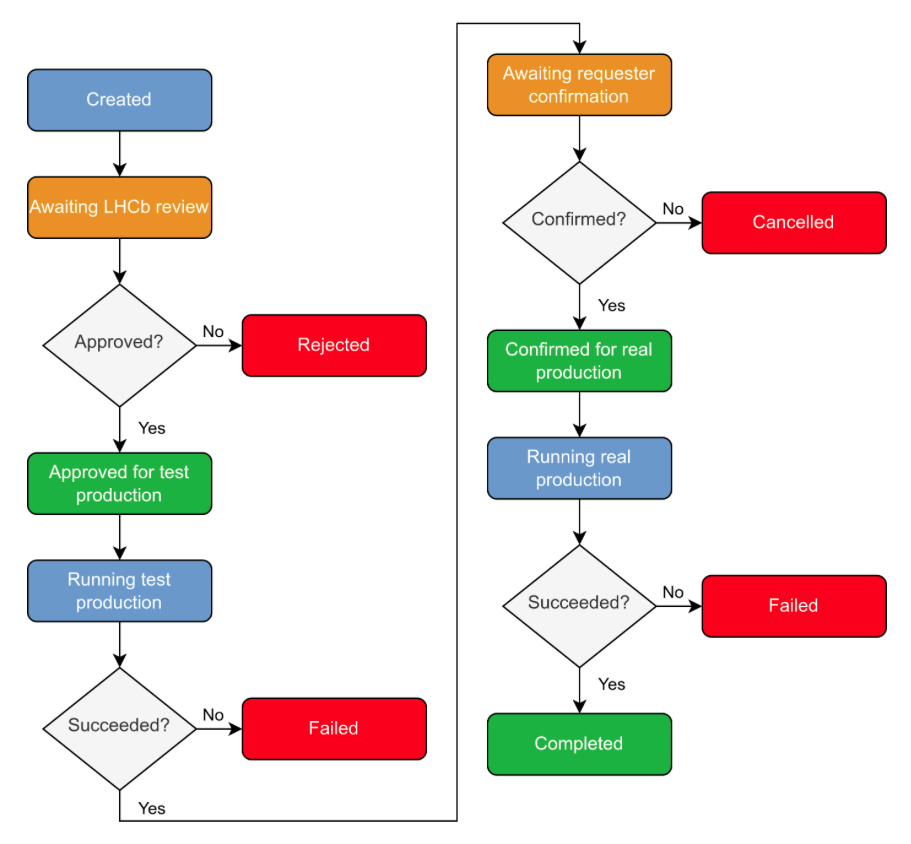}
    \caption{The LHCb Ntupling Service request container uses a carefully designed label system to streamline the user request. The rounded rectangles correspond to the various GitLab labels in the request repository. A query requires a review and multiple manual interventions for safety purposes.}
    \label{fig:label_workflow}
\end{figure}

Further streamlining is achieved by implementing a system using GitLab labels, indicating the status of the user request. The LHCb Ntupling Service has multiple triggers in place to manage the request based on the status represented in the issue label. The workflow from creating to completing a user request in terms of labels is detailed in~\Cref{fig:label_workflow}. The user is able to monitor status updates through the web application and is additionally notified via email whenever the status of the request changes.
Importantly, whenever a request is \textit{approved for test production}, an action is triggered that creates a draft merge request in the Analysis Productions Gitlab repository, which initiates a small test over a subset of data files, and from its output the Ntupling Service extracts the expected size and variables of the final Ntuple(s). Once the user inspects the test output and gives their green light, the LHCb Ntupling Service backend will transition the production request out of its draft status. At this stage, the Analysis Production manager must review the request once again and approve it to proceed into production where the full set of specified data is processed.

At this point, the backend has created a designated destination on CERN's \textit{EOSPUBLIC} storage space~\cite{eos} for the produced Ntuples, where they are transferred upon completion of the production. Once the label changes to \textit{completed}, the user is informed and is now able to download the data, as shown in~\Cref{fig:download_data}. 

\subsection{Security Considerations}
The LHCb Ntupling Service incorporates several measures to ensure the security of the system. A key priority is to prevent the execution of any malicious code within the CERN infrastructure, protecting both LHCb computing resources and the CERN WLCG environment. First, the DaVinci configuration files are generated only when the Ntupling Service backend receives the request, preventing them from being sourced from the browser and potentially modified there. Furthermore, a validation check is implemented to guarantee that the request does not handle data that has not been publicly released. This includes verifying restricted bookkeeping paths, Stripping lines, and data taking years. Additionally, each request undergoes multiple human interventions including a thorough review of the request and its configuration files. 
\label{sec:security}
\section{Conclusion and Outlook}
\label{Conclusion}
In this work, the LHCb Ntupling Service was presented for the first time. It provides a novel approach to on-demand publishing of LHCb data on the CERN Open Data Portal. The release format changed to the prior Run 1 release due to storage constraints arising from the sheer amount of data recorded during Run 2, leading to a novel and user friendly approach to open data. The LHCb Ntuple Wizard is used to generate production configuration files that can be read by Analysis Productions. The LHCb Ntupling Service embeds the Ntuple Wizard, manages incoming user requests and their status, and provides a link to the internal LHCb production system by spawning an on-demand production for the requested Ntuples. These Ntuples are then exposed back to the user for download, eliminating the need to execute complex experiment-specific software on the user side. 
Furthermore, multiple security measures are put in place to ensure the integrity of CERN computing resources. 

The LHCb Ntupling Service has undergone an internal testing phase, where valuable feedback was collected. Additionally, external theorists and phenomenologists were invited to test the service providing insights from an external perspective. Following the successful completion of these testing periods, future developments aim to integrate on-demand filtering of data samples and the inclusion of simulated samples. Furthermore, the user experience will be improved by documenting the nomenclature of LHCb tools and variables using Knowledge Graphs~\cite{KG}. Finally, there are plans to promote certain requests to permanent records on the CERN Open Data Portal with corresponding DOIs, facilitating reusibility and simplifying the process of citing the data.

\end{document}